\documentclass[12pt,preprint]{aastex}

\slugcomment{}

\shorttitle{ABUNDANCES OF HD~88609}
\shortauthors{HONDA et al.}

\begin{document}

\title{Neutron-capture elements in the very metal-poor star HD~88609: another star with excesses of light neutron-capture elements\altaffilmark{1}}

\author{Satoshi Honda\altaffilmark{2,3}, Wako Aoki\altaffilmark{2}, Yuhri Ishimaru\altaffilmark{4}, and Shinya Wanajo\altaffilmark{5}}

\altaffiltext{2}{National Astronomical Observatory, Mitaka, Tokyo,
181-8588, Japan; e-mail: honda@optik.mtk.nao.ac.jp, aoki.wako@nao.ac.jp}
\altaffiltext{3}{Current address: Gunma Astronomical Observatory, 6860-86 Nakayama, Takayama, Agatsuma, Gunma 377-0702, Japan}
\altaffiltext{4}{Academic Support Center, Kogakuin University, Hachioji, Tokyo 192-0015, Japan; kt13121@ns.kogakuin.ac.jp}
\altaffiltext{5}{Department of Astronomy, School of Science, University of Tokyo, Bunkyo-ku, Tokyo 113-8654, Japan; wanajo@astron.s.u-tokyo.ac.jp}
\altaffiltext{1}{Based on data collected at the Subaru Telescope,
which is operated by the National Astronomical Observatory of Japan.}

\begin{abstract}
We obtained a high resolution, high signal-to-noise UV-blue spectrum
of the extremely metal-poor red giant HD~88609 to determine the
abundances of heavy elements. Nineteen neutron-capture elements are
detected in the spectrum. Our analysis revealed that this object has
large excesses of light neutron-capture elements while heavy
neutron-capture elements are deficient.  The abundance pattern shows a
continuously decreasing trend, as a function of atomic number, from
Sr to Yb, which is quite different from those in stars with excesses
of r-process elements. Such an abundance pattern is very similar to that
of HD~122563 that was studied by our previous work. The results
indicate that the abundance pattern found in the two stars could
represent the pattern produced by the nucleosynthesis process that
provided light neutron-capture elements in the very early Galaxy.

\end{abstract}


\section{Introduction}

Over the past few decades, a considerable number of studies have been
conducted to identify the origin of neutron-capture
elements. Measurements of heavy elements for very metal-poor stars
provide a unique opportunity to determine the abundance patterns
produced by individual nucleosynthesis processes in the early Galaxy
\citep[e.g., ][]{mcwilliam95b}, which are key to understanding the
astrophysical sites of the processes. In particular, detailed
abundance studies for selected neutron-capture enhanced, metal-poor
stars have revealed that heavy neutron-capture elements ($Z\geq 56$)
show an abundance pattern that is identical, within the measurement
errors, to that of the r-process component of solar-system material
\citep[e.g., ][]{sneden96}.

However, these objects show some deficiencies in light neutron-capture elements ($Z < 56$) compared to the scaled solar r-process curve \citep[e.g.,][]{sneden00,hill02}.
This means that the r-process abundance pattern in solar-system material is not fully explained by the component that makes the pattern found in such r-process-enhanced stars (sometimes referred to as ``main'' r-process), but another component that yields light neutron-capture elements with small production of heavy ones is required.
Indeed, a some fraction of very metal-poor stars have high
abundance ratios of light neutron-capture elements Sr, Y, and Zr,
while their heavy ones (e.g., Ba, Eu) are very deficient
\citep[e.g.,][]{mcwilliam98,aoki05}. 
Such objects possibly record the abundance patterns
produced by the second component.

The component that is responsible for light neutron-capture elements of the r-process fraction of 
solar-system material is not well understood, and no established name is given.
We here refer to this process as ``weak'' r-process \citep{wanajo06}.
It should be noted that the r-process component in the solar system is derived by subtracting the s-process component from the solar abundances, based on the s-process models \citep{kap89}.
Therefore, the weak r-process referred in this paper possibly corresponds to some processes other than r-process.

The bright metal-poor ([Fe/H]$=-2.7$) star HD~122563 is one of the stars having 
excesses of light neutron-capture elements. 
Our previous work obtained high resolution spectroscopy for
this object and determined the abundances of 19 neutron-capture
elements from Sr to Yb ($Z=38-70$) \citep[][; hereafter Paper
I]{honda06}. The abundances of neutron-capture elements in this star
continuously decrease with the increase of atomic number. That is, the
elements with intermediate mass ($41 \leq Z \leq 47$) have moderate
enhancements with respect to heavy ones ($Z\geq 56$). 
This is the first determination of the overall abundance pattern that could 
represent the yields of the weak r-process mentioned above.

In this paper, we report the abundance pattern of neutron-capture
elements in the extremely metal-poor red giant HD~88609. Previous
studies \citep[e.g., ][]{honda04} showed that this object has large
excesses of light neutron-capture elements and underabundances of
heavy ones similar to HD~122563. The metallicity of HD~88609 ([Fe/H]$\sim
-3.0$) is even lower than that of HD~122563 ([Fe/H]$\sim -2.7$). Hence,
HD~88609 could be an ideal object in order to study the nature of 
the weak r-process. 
However, abundance measurements of the neutron-capture elements for this 
object are quite limited because of the deficiency of heavy neutron-capture elements. 
We determined the detailed
abundance pattern of heavy elements for this object based on
a high resolution, high signal-to-noise ratio spectrum. The
observation and abundance measurements are reported in \S~2 and 3,
respectively. 
The results are compared with previous work in \S~4. 
In \S~5, the results are compared with the abundance pattern of the solar-system
r-process component, and with that of HD~122563. 
Implications of our new measurement for HD~88609 are also discussed.

\section{Observations}

High dispersion spectroscopy of HD~88609 was carried out with the Subaru Telescope High Dispersion Spectrograph \citep[HDS: ][]{nogu02} on 30 April, 2004 and 19-20 October 2005.
Our spectrum covers the wavelength range from 3020 to 4780 {\AA} with a resolving power of $R$ = 90,000, as in the study for HD~122563 (Paper I).
The total exposure time of the two observing runs is 12600 seconds.
In addition, the spectrum around 5000 {\AA} obtained by \citet{honda04} was used to determine the abundances of Cu, Zn, Y, and Ba abundances (see Table 1 for the details of the observations).

The data reduction was carried out in a standard manner using the IRAF echelle package\footnote{IRAF is distributed by National Optical Astronomy Observatories, which are operated by the Association of Universities for Research in Astronomy, Inc., under cooperative agreement with the National Science Foundation} as described in Paper I and \citet{honda04a}. 
The signal to noise ratio (S/N) of the spectrum (per 0.9 km s$^{-1}$ pixel) estimated from the photon counts is 50 at 3100 {\AA}; 200 at 3500 {\AA}; 370 at 3900 {\AA}; 510 at 4200 {\AA} and 650 at 4500 {\AA}.
These S/N ratios are not as high as those of the spectrum of HD~122563 (Paper I), but sufficient for our purposes.

\section{Analysis}

For our quantitative abundance measurements, we used the analysis program SPTOOL developed by Y. Takeda (private communication), based on Kurucz's ATLAS9/WIDTH9 \citep{kurucz93}.  
SPTOOL calculates synthetic spectra and equivalent widths of lines on the basis of the given model atmosphere, line data, and chemical composition, under the assumption of LTE.

We adopted the model atmosphere parameters (effective temperature: $T_{\rm eff}$, gravity: $g$, micro-turbulent velocity:  $v_{\rm turb}$, and metallicity: [Fe/H]) derived by \citet{honda04}: $T_{\rm eff}$ = 4550 K, $\log g$ = 1.1,  $v_{\rm turb}$ = 2.4 km s$^{-1}$, and [Fe/H] $= -3.07$.

\begin{figure}[p]
\includegraphics[width=10cm,angle=-90]{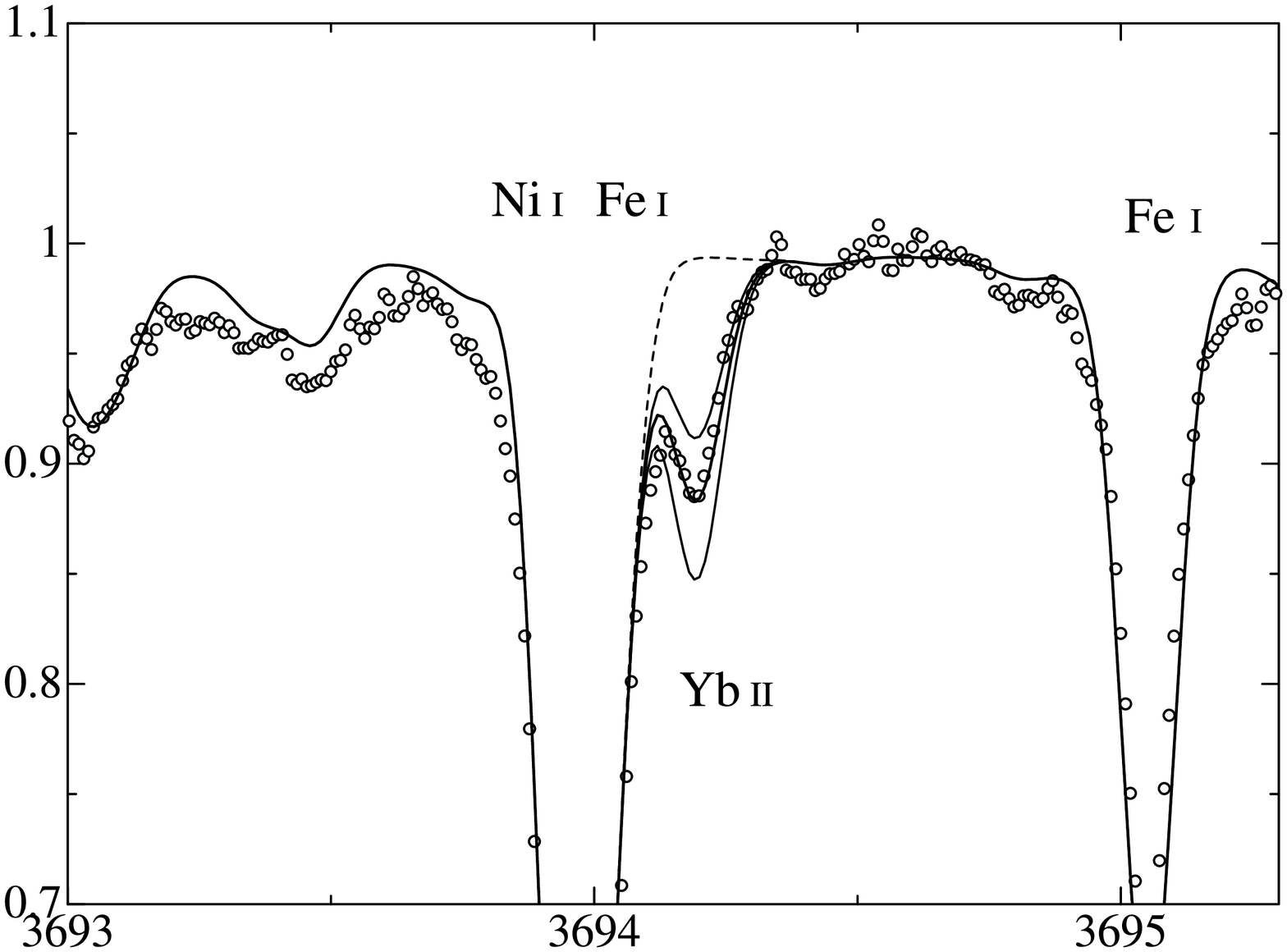}
\includegraphics[width=10.5cm,angle=-90]{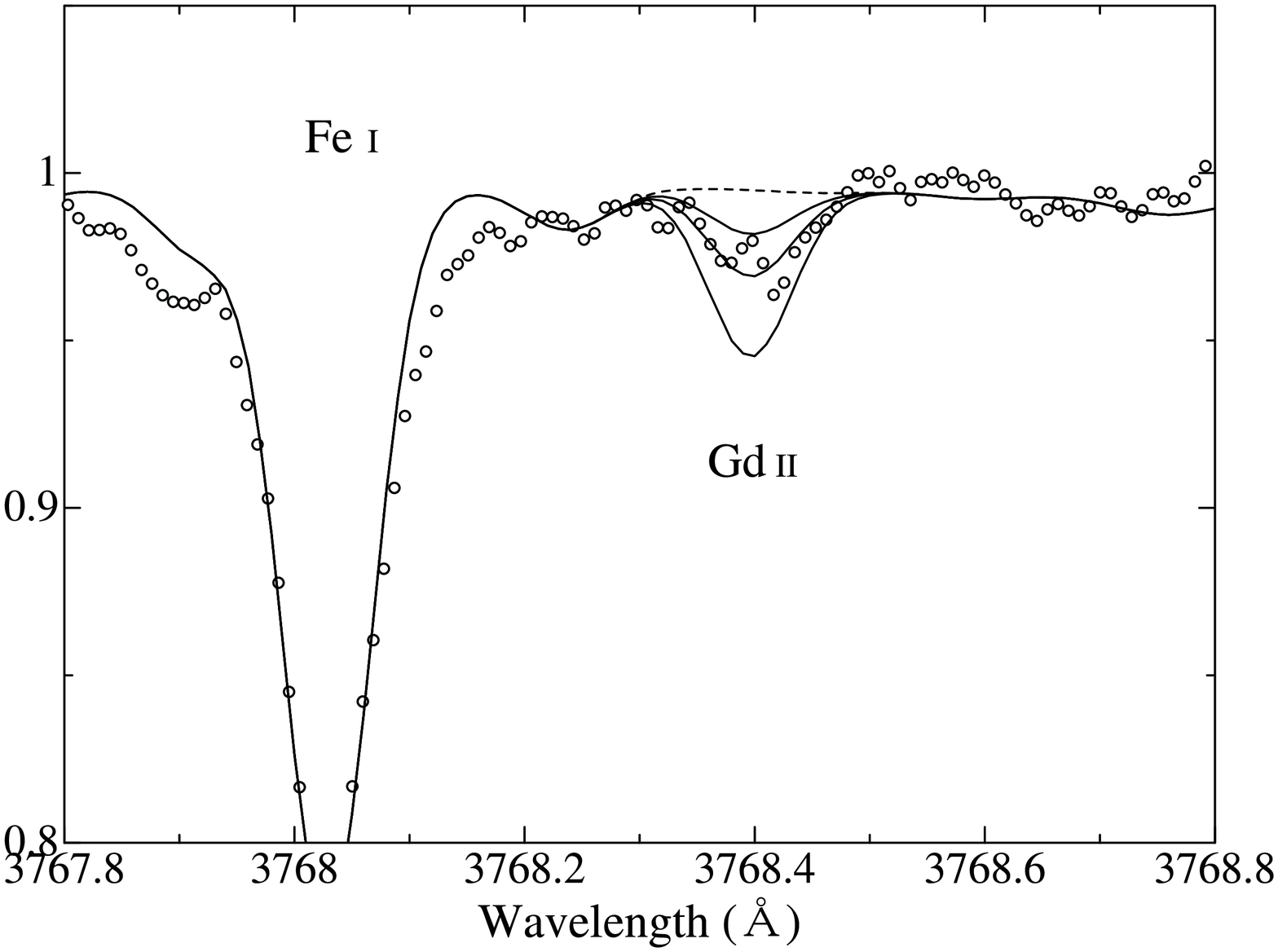}
\caption{The spectra of \ion{Yb}{2} 3694 {\AA}, and \ion{Gd}{2} 3768 {\AA}.
Dots: observations; 
solid lines: synthetic spectra computed for the adopted abundance (see Table 1) 
and values $\pm$0.3 dex different; 
dashed lines: synthetic spectra with no contribution from the line of interest.}
\label{fig:fig1}
\end{figure}

We adopted the line list used in the previous study for HD~122563 (Paper I), which is given in Table \ref{tab:table2}.
All the lines used in the previous study were also detected in the spectrum of HD~88609.
An advantage in the present analysis is that, since the metallicity of this object is lower than that of HD~122563, the effect of the blends of lines decreases.
Equivalent widths of isolated lines are measured by fitting a gaussian profile.
The lines of \ion{Cu}{1} 5105, \ion{Zn}{1} 4810, \ion{Y}{2} 4883, \ion{Y}{2} 5087, and \ion{Ba}{2} 4934 are analyzed using the spectrum of \citet{honda04a}, which has sufficient S/N in this wavelength region (150 at 5000 {\AA}).
The values of the equivalent widths of \ion{Ba}{2} 5853 and 6141 {\AA} are adopted from \citet{johnson02a}.
The results are given in Table \ref{tab:table3}.
A spectrum synthesis technique is applied to lines which are severely affected by blending and/or hyperfine splitting.
Equivalent widths of such lines given in Table \ref{tab:table2} are calculated by spectrum synthesis for the derived abundances.
We used the solar system abundances obtained by \citet{asplund05} to derive [X/Fe] values.
Figure \ref{fig:fig1} depicts representative examples of the quality of the observed spectra, and the fits to synthetic spectra.
The abundances of HD~122563 are also given in this table for comparison purposes.
We note that the list of abundances in Table 2 of Paper I contains small mistakes, which are corrected in Table 3 of the present paper.
The difference is at most 0.08 dex (Ce), and does not have significant influence on the conclusions of the previous paper.

The size of the random errors is estimated from the standard deviation (1$\sigma$) of the abundances derived from individual lines for elements that have four or more lines available for the abundance analysis.
For the abundances of elements based on less than four lines, we employ the mean of the random errors estimated from those elements with four or more lines available (0.12 dex).
For most of these elements, the abundances derived from individual 
lines distribute within 0.12 dex around the mean abundance that is 
adopted as the final result.
Exceptions are Cu, Dy, and Er, for which we adopt 0.20 dex as the 
random errors.
We also estimate the random errors in the abundance measurement from the uncertainties of equivalent width measurements, which are estimated using the relation $\sigma_{\rm w}\simeq({\lambda}n_{\rm pix}^{1/2})/(R[S/N])$ \citep{norris01}.
The errors of abundances from equivalent width measurement are smaller than 0.12 dex for most lines.
The estimated errors for \ion{Cu}{1} {$\lambda$}5105 and \ion{Gd}{2} ${\lambda}$3549 are very large, 
because of the weakness of these lines.
However, the abundance results do not significantly change if these lines are excluded in the analysis.
We note that, since contaminations to the \ion{Eu}{2} $\lambda$3819, $\lambda$4205, and \ion{Gd}{2} $\lambda$3481 are possibly significant, these lines are not used to derive final results.

Although the S/N ratio of the HD~122563 spectrum is better than the HD~88609 one, 
the standard deviation of the abundances derived from four or more lines for HD~88609 (0.12 dex) is slightly smaller than those for HD~122563 (0.14 dex).
This suggests that factors other than the quality of the spectrum also affect the random errors.

The atmospheric parameters of HD~88609 are quite similar to those of HD~122563, and the same line set is used for the abundance analysis.
Therefore, we adopt the errors due to the uncertainties of the atmospheric parameters 
($\Delta T_{\rm eff}$ = 100 K, $\Delta \log g$ = 0.3 dex, $\Delta v_{\rm turb}$ = 0.5 km s$^{-1}$, and $\Delta$[Fe/H] = 0.20 dex) 
estimated for HD~122563 (Table 3 of Paper I).
The total systematic errors which are shown in Table 3 ($\sigma$) are derived by the root sum square (r.s.s.) of the uncertainties contributed by the four parameters.
Such systematic errors do not significantly affect the abundance ratios of neutron-capture elements, in particular those derived from species with the same ionization stage.
The effects of uncertainties of atmospheric parameters on abundance ratios (and the abundance pattern) of neutron-capture elements are estimated by the following procedure; 1) we calculate the average of the abundance changes due to the change of each atmospheric parameters (e.g., $\Delta T_{\rm eff}$) for neutron capture elements (19 species); 2) we calculate the deviation of the abundance changes for individual species from the above average value; 3) we adopt the r.s.s. of the above deviations for the four parameters as the systematic errors for the discussion on the abundance ratios ($\sigma_{\rm n-cap}$ in Table 3).
The typical value is 0.07 dex for the ionized species, while that is 0.12 dex for the neutral ones.

The radial velocity is measured using 14 clean iron lines.
The derived radial velocity is --37.69 $\pm$ 0.3~km s$^{-1}$ from the data of Oct. 2005.
This value agrees well with the value --37.28 $\pm$ 0.43~km s$^{-1}$ derived by our previous study  \citep{honda04a}.
There is, so far, no evidence that this object belongs to a binary system.

\section{Comparisons with previous studies}

\begin{figure}[p]
\includegraphics[width=6.4cm,angle=-90]{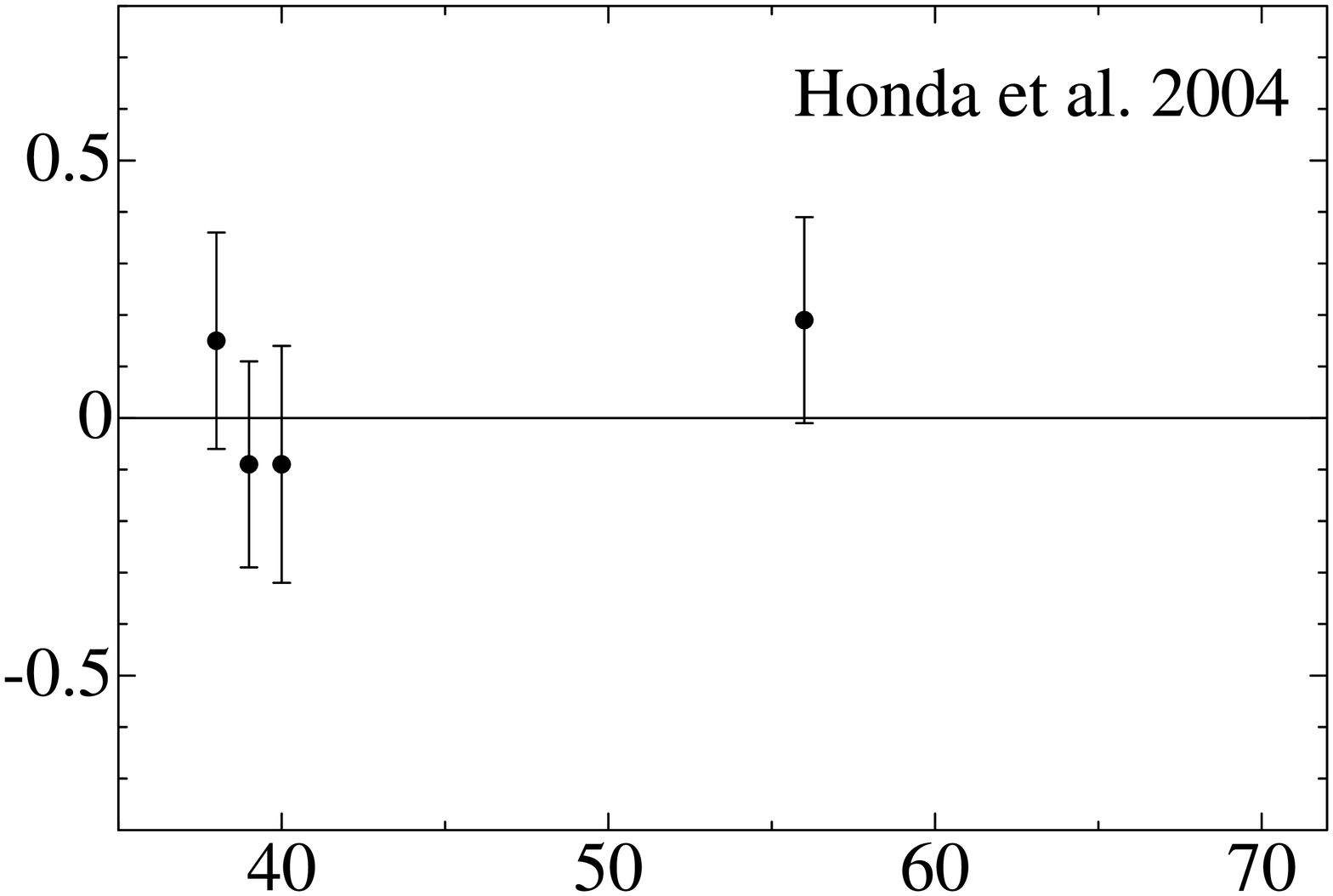}

\includegraphics[width=6.4cm,angle=-90]{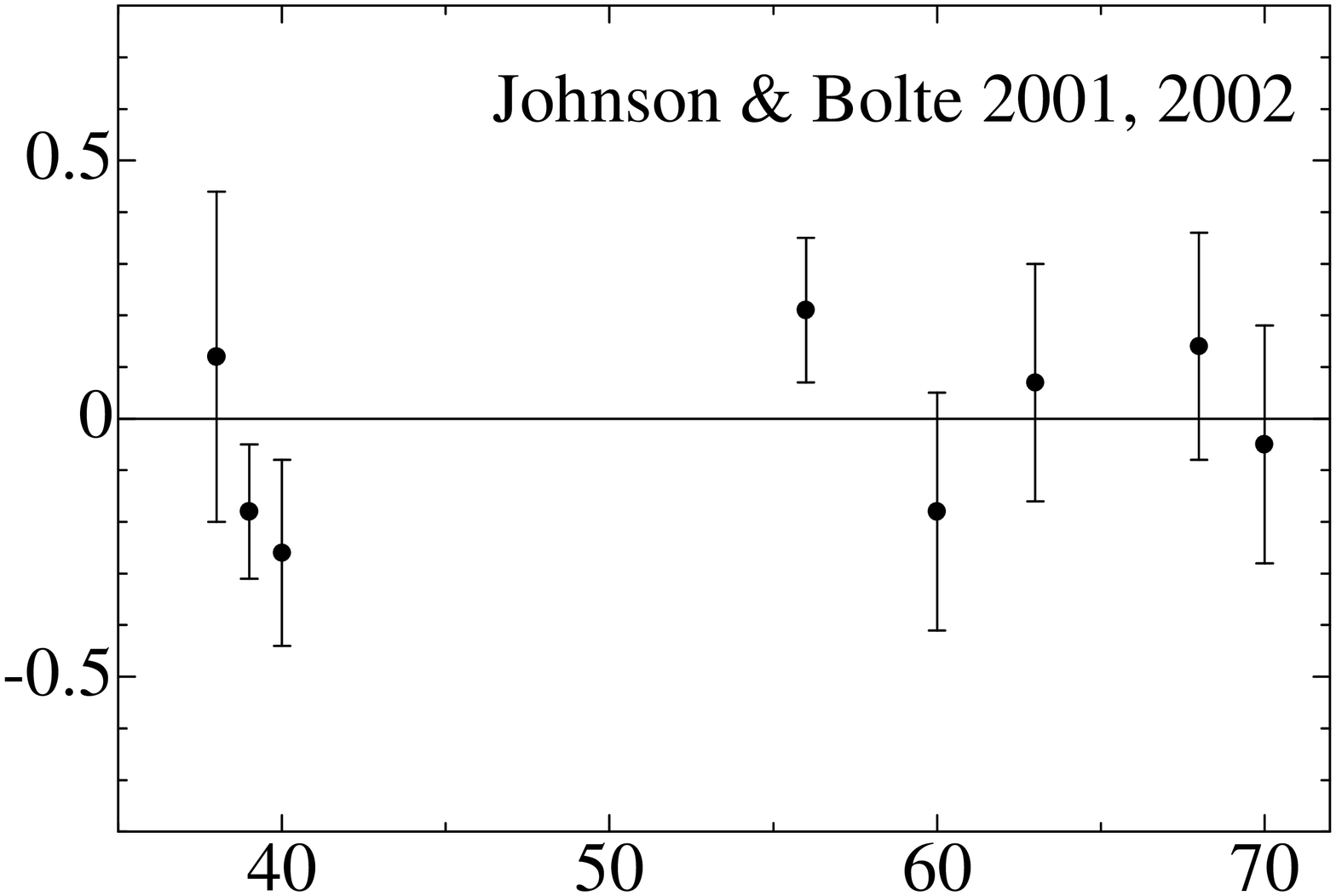}

\includegraphics[width=7cm,angle=-90]{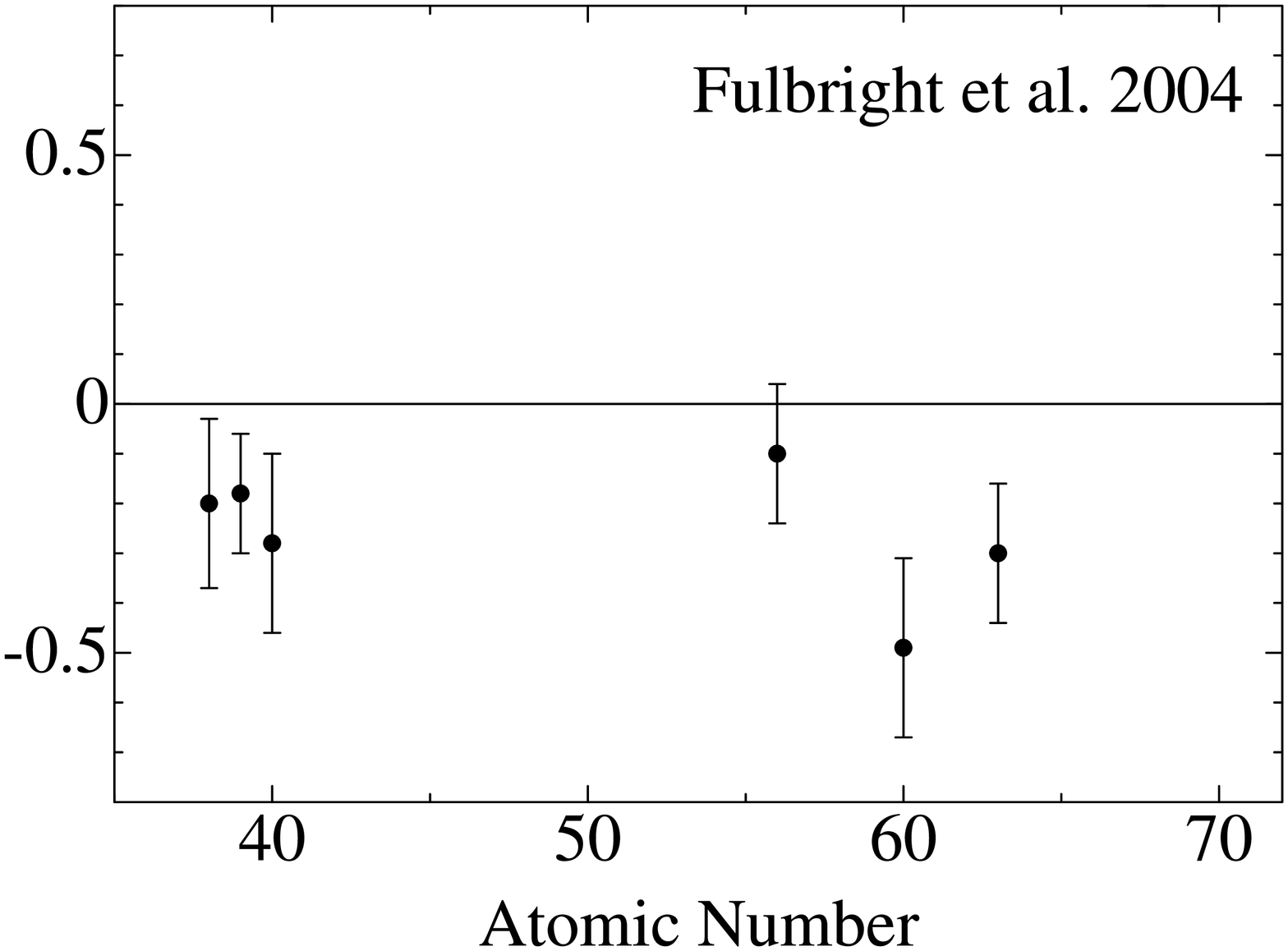}
\caption{Comparisons with the abundances derived by the present analysis and 
previous studies (Honda et al. 2004, Johnson \& Bolte 2001, 2002, and Fulbright et al. 2004) in the sense
log$\varepsilon_{\rm this}$ $-$ log$\varepsilon_{\rm other}$, 
as a function of atomic number. }
\label{fig:fig2}
\end{figure}

In spite of the brightness of HD~88609, abundance measurements of neutron-capture elements have been reported only for several species in previous work, because of the deficiency of heavy neutron-capture elements.
Figure \ref{fig:fig2} shows comparisons of the abundances of neutron-capture
elements derived by the present analysis with recent studies by \citet{johnson02a}, \citet{fulbright04}, and \citet{honda04}.

Since the present work adopts the atmospheric parameters determined by \citet{honda04} for the abundance analysis, no systematic error is expected.
Although Sr, Y, Zr, and Ba are detected by \citet{honda04}, many lines are added to previous 
data by this analysis.
The abundances of these elements determined by the two studies agree within the errors.
We note that the Ba abundance derived by the present work is higher than the value of \citet{honda04}, but this is because the present analysis includes the two lines in the red region that give systematically higher abundances than those from the two resonance lines.

\begin{figure}[p]
\includegraphics[width=12cm]{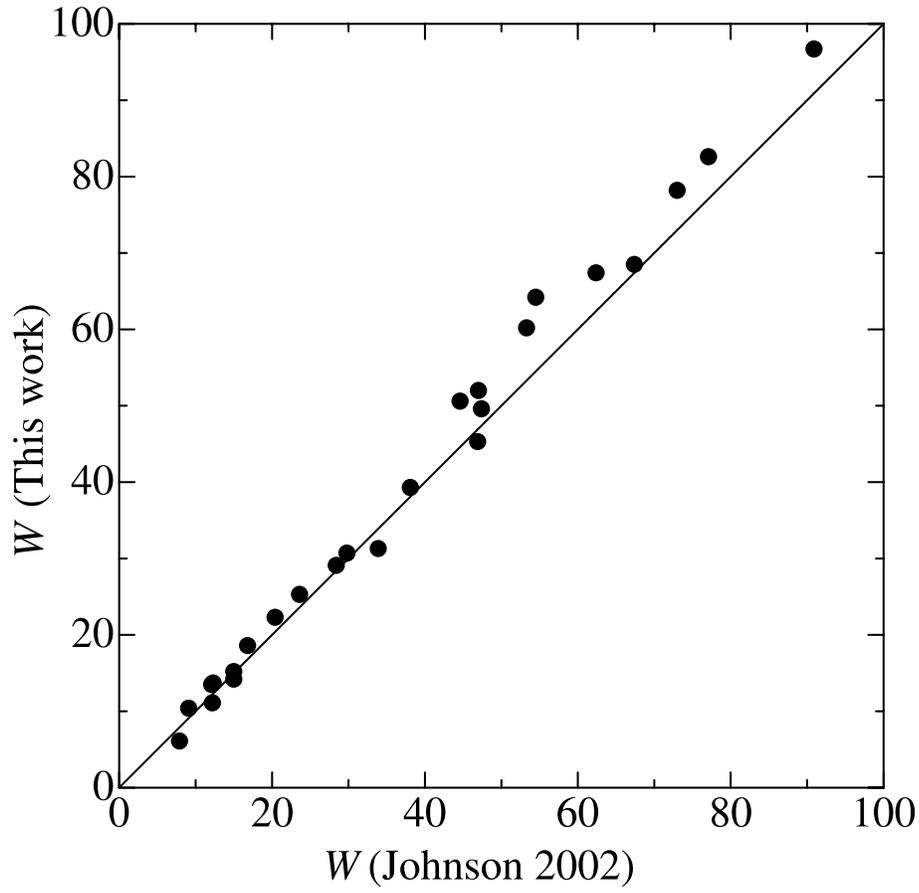}
\caption{Comparison of equivalent width ({\it W}) measurements (m{\AA}) by Johnson (2002) and this work.}
\label{fig:fig2a}
\end{figure}

\citet{johnson02a} and \citet{johnson01,johnson02} analyzed the neutron-capture elements in this object.
A comparison of equivalent widths of 25 lines of Zn, Y, Zr, and Ba (Figure \ref{fig:fig2a}) shows a systematic difference by 11\% between the two studies.
This results in the abundance difference of at most 0.1 dex.
Indeed, the abundances of Y, Zr, and Ba derived by our analysis technique using the equivalent width of \citet{johnson02a} and \citet{johnson01,johnson02}, differ only 0.03 -- 0.08 dex from their results.

The abundances determined by the present study agree well with their results.
However, the atmospheric parameters adopted in their analysis differ from ours:
the effective temperature adopted in our analysis is 150K higher than that of \citet{johnson02a}, while our gravity ($\log g$) is 0.7 dex higher than their value.
The differences of these parameters could result in systematic abundance differences of about 0.3 dex, which are not found between the two studies.

\citet{fulbright04} reanalyzed the spectrum of HD~88609 partially using the equivalent widths of \citet{johnson02a}.
Since they derived the atmospheric parameters by a similar method to \citet{honda04}, the values are in good agreement with ours.
However, systematic difference in the derived abundances is found between the two works (Figure \ref{fig:fig2}).

The comparisons of our results with those of \citet{johnson02a}, \citet{johnson01,johnson02}, and \citet{fulbright04} show that systematic differences exist in abundances of neutron-capture elements, taking account of adopted atmospheric parameters.
A possible reason for this discrepancy is the difference of abundance analysis tools.
We note that, although we can not identify the reason for this systematic difference, the abundance ratios (patterns) of neutron-capture elements show relatively good agreement between these works.


\section{Discussion and concluding remarks}

Our abundance analyses for HD~88609 based on high resolution
spectroscopy determined the abundances of 19 neutron-capture elements
as for HD~122563 by our previous work.  The discussion in this section
focuses on the comparison of the abundance patterns of these elements
between the two objects, and implications of the new results obtained
for HD~88609.

\begin{figure}[p]
\includegraphics[width=12cm,angle=-90]{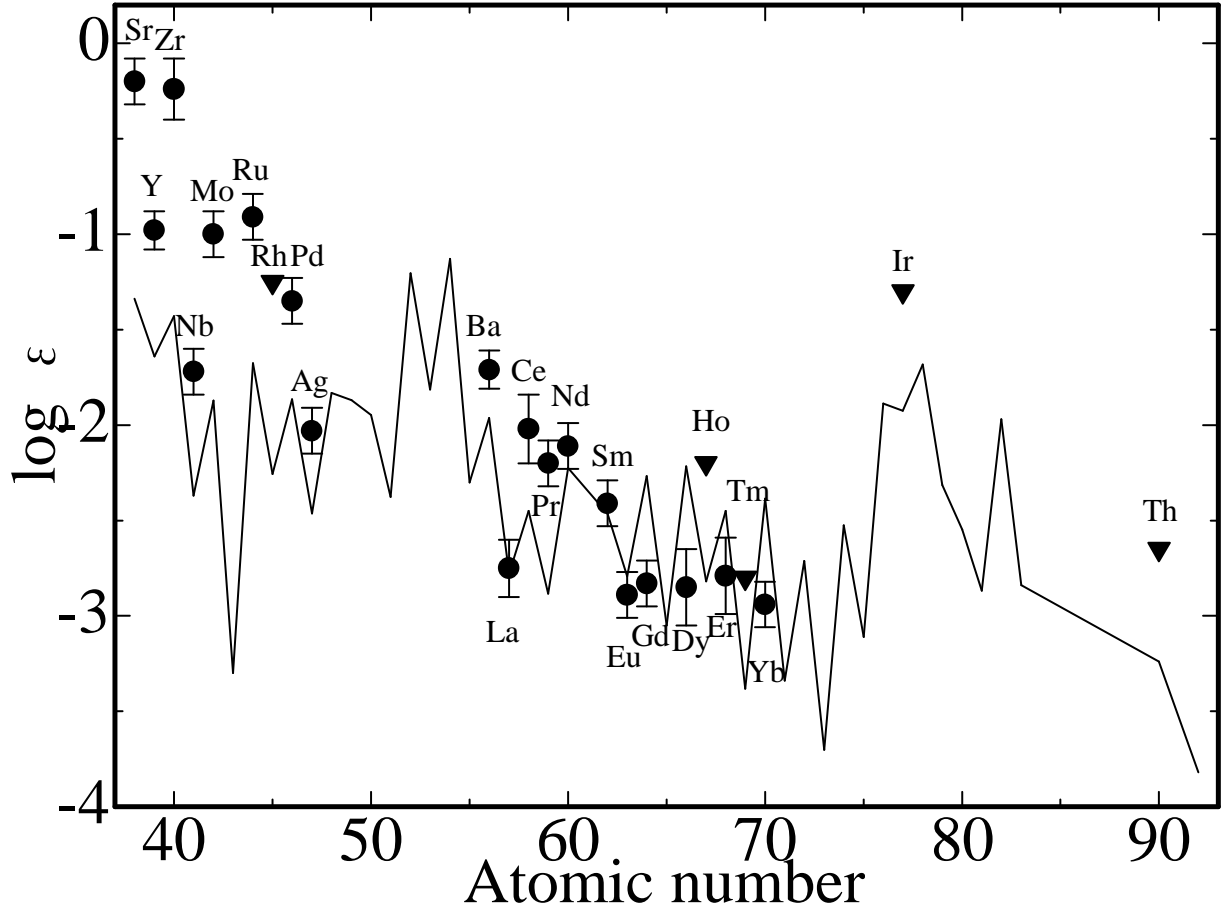}
\caption{The abundances of HD~88609 compared to the scaled solar-system r-process pattern (normalised at Eu). The solar system abundances are taken from \citet{asplund05}, and the r-process fraction in the solar system given by Simmerer et al. (2004).}
\label{fig:fig3}
\end{figure}

Figure \ref{fig:fig3} shows the abundance pattern of HD~88609, compared
with that of the solar-system r-process component.  This figure
clearly shows that the abundances of light neutron-capture elements
(38 $\leq$ Z $\leq$ 47) are much higher than those of heavy ones in
HD~88609, compared to the solar-system r-process pattern.  This
indicates that the neutron-capture elements of HD~88609, as
we found for HD~122563, are not a result of the main r-process, but were
produced by another process.

\begin{figure}[p]
\includegraphics[width=12cm,angle=-90]{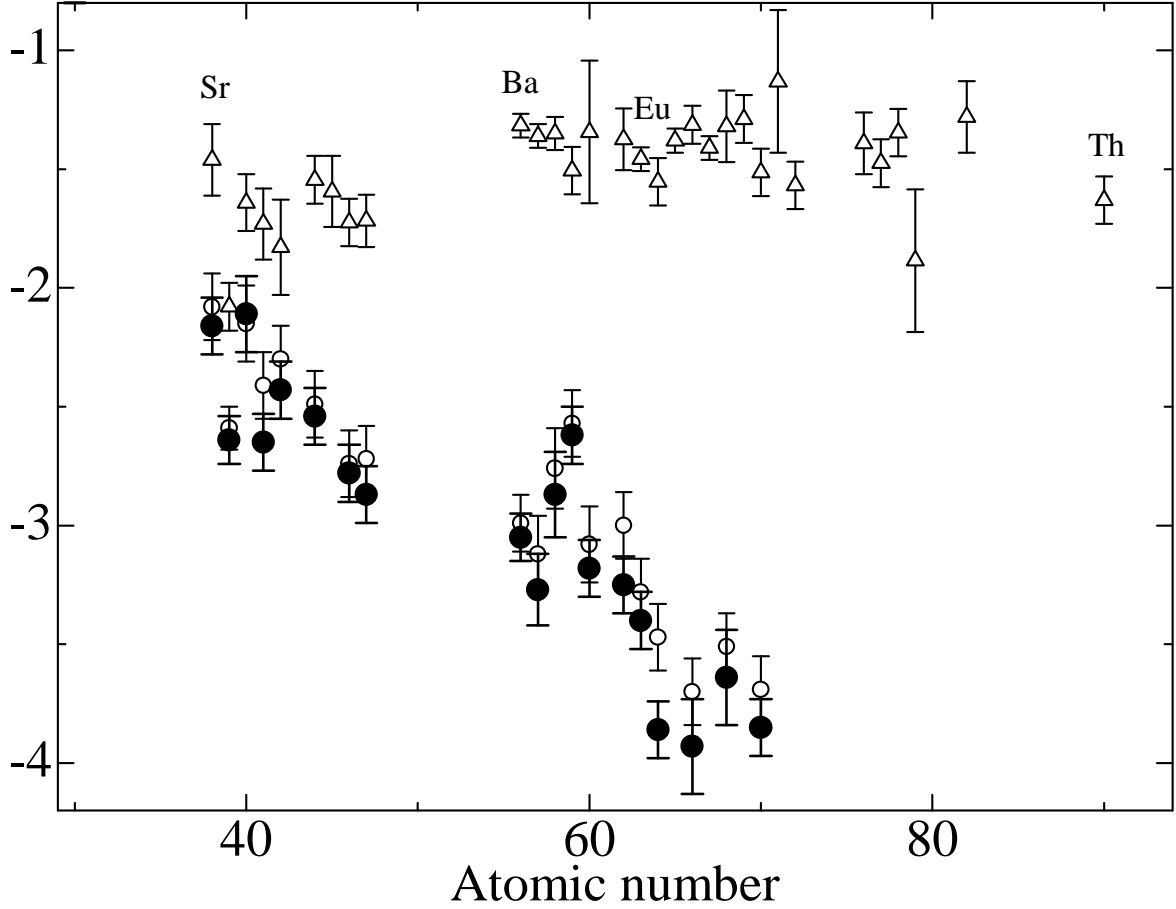}
\caption{Logarithmic differences from the solar system r-process 
pattern (log$\varepsilon_{\rm object}$ $-$ log$\varepsilon_{\rm solar-r}$). The open triangles mean CS~22892-052, the open circles mean HD~122563, and the filled circles mean HD~88609.}
\label{fig:fig4}
\end{figure}

Figure \ref{fig:fig4} shows the abundance differences from the solar system
r-process component in the logarithmic scale for HD~88609 (this work), HD~122563 (Paper I),
and CS~22892-052 \citep{sneden03}.
A clear result is that the abundance pattern of HD~88609 is very similar to that of HD~122563.
This result was partially expected from the sample selection: we selected objects
having high abundance ratios between light and heavy neutron-capture
elements (e.g., Sr/Ba). However, an important finding is that the two
objects have quite similar abundance patterns of elements between the
first and second abundance peaks ($41 \leq Z \leq 47$), which are
determined for HD~88609, for the first time, by the present work.

We discussed possible nucleosynthesis processes that explain the abundance pattern of HD~122563 in Paper I. 
The main r-process and the main s-process are excluded, because they produce much higher
abundance ratios of heavy to light neutron-capture elements than found
in HD~122563. The weak s-process is unlikely to be the source of
neutron-capture elements of HD~122563, because of the moderate
excesses of elements with intermediate mass (e.g., Pd). 
We concluded that an unknown component of the r-process is responsible for the light
neutron-capture elements in HD~122563. We referred to the ``lighter
element primary process (LEPP)''  proposed by \citet{travaglio04} and the
``weak'' (or ``failed'') r-process \citep{wanajo06} as the possible
sources of neutron-capture elements in Paper I.
Our present study reveals that HD~88609 is the second example of
metal-poor stars that have the abundance pattern representing such a
process, and HD~122563 is not a peculiar object.

\begin{figure}[p]
\includegraphics[width=12cm,angle=-90]{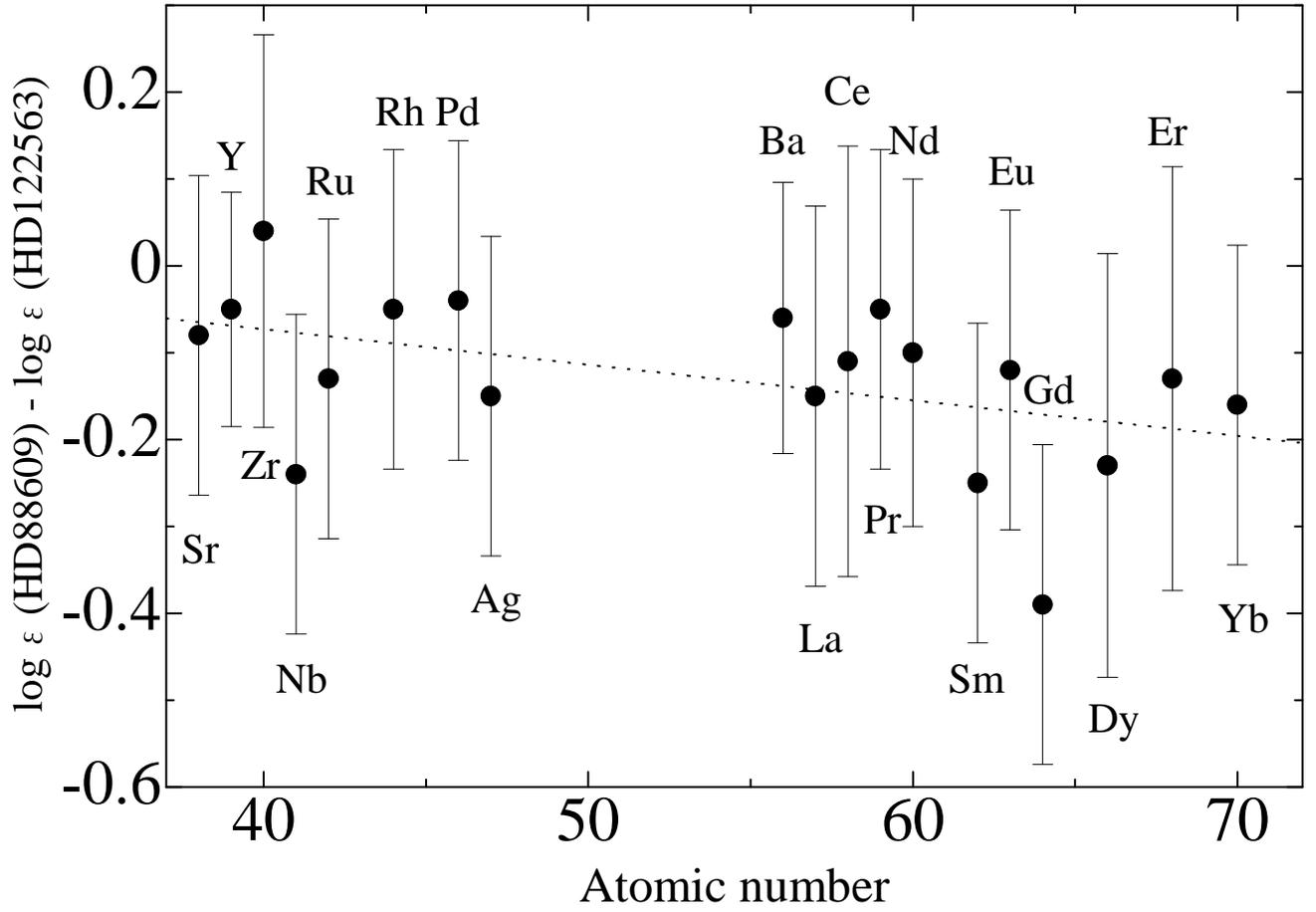}
\caption{Comparison of the elemental abundances between HD~88609 and HD~122563 (log$\varepsilon_{\rm HD88609}$ $-$ log$\varepsilon_{\rm HD122563}$), as a function of atomic number.}
\label{fig:fig5}
\end{figure}

Here we inspect the difference of the abundance patterns between the
two objects in some detail.  
Figure \ref{fig:fig5} shows the abundance differences
of neutron-capture elements on a logarithmic scale between HD~88609 and
HD~122563 as a function of atomic number. The least square fit to the
data points suggests that the decreasing trend is steeper in HD~88609
than that in HD~122563. The regression analysis and the Spearman rank
correlation test indicate that the null hypothesis that there is no
correlation between the abundance difference and the atomic number is
rejected at the 95\% confidence level. Hence, though the result is not
definitive, a correlation between the two values likely exists.

Our conclusion here is that the abundance patterns of neutron-capture elements in HD~88609 and HD~122563 are quite similar, though a small difference is suggested. 
One might recall the proposed uniqueness of the abundance pattern produced by the main r-process, which is confirmed by the observations of several r-process enhanced, extremely metal-poor stars \citep[cf.][]{cowan06}. 
Theoretical studies based on the neutrino wind scenario also predict the robustness of abundance patterns when the requisite physical conditions (e.g., high entropy, short dynamic timescale, or low electron fraction) for the production of the main r-process nuclei are obtained \citep{Wana01, Wana02}. 
However, the similarity of the abundance patterns found in the two objects can not be interpreted as uniqueness of the pattern produced by the weak r-process (or LEPP), because the two objects were selected to have similarly high Sr/Ba abundance ratios. 
Model calculations for insufficient conditions for the main r-process show that the r-processing appears to be highly sensitive to the model parameters (e.g., electron fraction), resulting in
various abundance curves (S. Wanajo et al., in preparation). 
It may be possible, therefore, that we have just found two similar ones out of various abundance curves from the very weak r-process (producing only Sr, Y, and Zr) to the main r-process (producing Ba, Eu, and heavier).
Further observations of objects having different Sr/Ba
ratios will uncover the diversity of abundance patterns.
In case that the similarity of the weak r-process patterns
is confirmed for a larger sample of stars, it will provide unique
constraints on the theoretical r-process modeling.

We also discussed the origin of heavy neutron-capture elements ($Z\geq
56$) in HD~122563, whose abundances significantly deviate from the solar-system
r-process abundance pattern, in Paper I. We argued the possibility of
some contamination of the s-process that can partially explain the
excesses of Ba, La, and Ce with respective to Eu, though an s-process contribution is unlikely significant at the metallicity of that object, i.e. [Fe/H]$=-2.7$. The
metallicity of HD~88609 ([Fe/H]$\sim -3.0$) is even lower than that of
HD~122563. No previous studies suggest contributions of the s-process
to objects with such low metallicity.\footnote{Exceptions are
carbon-enhanced metal-poor stars, which are presumably affected by
nucleosynthesis of asymptotic giant branch stars and mass transfer
across a binary system \citep[e.g.,][]{aoki07}. HD~88609, as well as HD~122563, is not an object belonging to this class.} Therefore, we conclude that the
deviation of the abundances of heavy neutron-capture elements in
HD~88609, and probably in HD~122563, from the main r-process pattern is not a result of contamination
of the s-process yields, but implies that the weak r-process produces a different abundance pattern of heavy
neutron-capture elements from that produced by the main r-process.
This result gives another constraint on the models of the process that
yielded light neutron-capture elements in the very early Galaxy.

\acknowledgments

We would like to thank Dr. Sean G. Ryan for his great help and useful comments.
We also thank Dr. Akito Tajitsu for their support to our observations.
This work was supported in part by a Grant-in-Aid for the Japan-France
Integrated Action Program (SAKURA), awarded by the Japan Society for the
Promotion of Science, and Scientific Research (17740108) from the
Ministry of Education, Culture, Sports, Science, and Technology of Japan.
Most of the data reduction was carried out at the Astronomical Data 
Analysis Center (ADAC) of the National Astronomical Observatory of Japan.

\clearpage

\begin{deluxetable}{lccc}
\tablewidth{0pt}
\tablecaption{OBSERVATION LOG AND ADOPTED ATMOSPHERIC PARAMETERS\label{tab:table1}}
\startdata
\tableline
Obs. Date & April 20. 2004 & October 19-20. 2005\\\hline
Wavelength & 3070 - 4780 {\AA} & 3020 - 4610 {\AA}\\
Resolution & 90,000 & 90,000\\
Exp. Time & 3600 sec & 9000 sec\\
SN @ 3500 & 120 & 150\\
SN @ 4000 & 240 & 330\\\hline
$T_{\rm eff}$ & 4550 K\\
$\log g$  & 1.1\\
$v_{\rm turb}$ & 2.4 km s$^{-1}$\\
$$[Fe/H] & --3.07 $\pm$ 0.2\\
\tableline
\enddata
\end{deluxetable}

\begin{deluxetable}{lcccccccc}
\tablewidth{0pt}
\tablecaption{LINE DATA AND EQUIVALENT WIDTHS \label{tab:table2}}
\startdata
\tableline
\tableline
Wavelength & L.E.P.(eV) & log$gf$ & log$\epsilon$ & {\it W}(m{\AA}) & $\Delta${\it W} & error & ref \\\hline
\ion{Cu}{1}, $Z=29$ &  &  &  &  &  \\\hline
3247.53 & 0.000  & --0.060  & 0.75  & 110.2* & 1.07 & 0.03 & 8\\
3273.95 & 0.000  & --0.360  & 0.85  & 103.2* & 0.99 & 0.03 & 8\\
5105.55 & 1.390  & --1.520  & 0.39  & 1.8**  & 1.16 & 0.49 & 2\\\hline
\ion{Zn}{1}, $Z=30$ &  &  &  &  &  \\\hline
3302.98 & 4.030  & --0.057  & 1.90  & 15.6* & 0.92 & 0.03 & 1 \\
3345.02 & 4.078  &   0.246  & 1.96  & 26.3 & 0.83 & 0.02 & 1 \\
4722.15 & 4.030  & --0.390  & 1.91  & 10.4  & 1.24 & 0.06 & 5 \\
4810.54 & 4.080 &  --0.170  & 1.90  & 14.2**& 1.22 & 0.04 & 5\\\hline
\ion{Sr}{1}, $Z=38$ &  &  &  &  &  \\\hline
4607.33 & 0.000  &   0.280  & --0.21  & 2.2 &  0.24 & 0.04 & 7\\\hline
\ion{Sr}{2}, $Z=38$ &  &  &  &  &  \\\hline
4077.71 & 0.000  &   0.170  & --0.20 & 169.8 & 0.31 & 0.01 & 6 \\
4215.52 & 0.000  & --0.170  & --0.18  & 154.2 & 0.28 & 0.01 & 6\\\hline
\ion{Y}{2}, $Z=39$ &  &  &  &  &  \\\hline
3327.88 & 0.410  &   0.130  & --0.93  & 55.4 & 0.86 &0.02 & 8 \\
3549.01 & 0.130  & --0.280  & --1.00  & 50.6  & 0.56 & 0.01 & 7\\
3584.52 & 0.100  & --0.410  & --1.09  & 41.5* & 0.53 & 0.01 &10 \\
3600.74 & 0.180  &   0.280  & --1.15  & 68.5 & 0.52 & 0.01 & 7 \\
3611.04 & 0.130  &   0.010  & --1.12  & 60.2 & 0.51 & 0.01 &  7\\
3628.70 & 0.130  & --0.710  & --0.96  & 31.3 & 0.50 & 0.01 & 10\\
3710.29 & 0.180  &   0.460  & --1.10  & 82.6 & 0.45 & 0.01 & 10\\
3747.55 & 0.100  & --0.910  & --0.97  & 25.3 & 0.43 & 0.01 & 7 \\
3774.33 & 0.130  &   0.210  & --1.04  & 78.2 & 0.42 & 0.01 & 6\\
3788.70 & 0.100  & --0.070  & --1.02  & 67.4 & 0.41 & 0.01 & 6\\
3818.34 & 0.130  & --0.980  & --0.77  & 29.7 & 0.40 & 0.01 & 6 \\
3950.36 & 0.100  & --0.490  & --0.95  & 49.6 & 0.35 & 0.01 & 6 \\
4398.01 & 0.130  & --1.000  & --0.83  & 29.1 & 0.26 & 0.01 & 6 \\
4883.69 & 1.080  &   0.070  & --0.91  & 22.3** & 1.20 & 0.03 & 6\\
5087.43 & 1.080  & --0.170  & --0.94  & 13.7** & 1.16 & 0.04 & 6\\\hline
\ion{Zr}{2}, $Z=40$ &  &  &  &  &  \\\hline
3438.23 & 0.090  &   0.420  & --0.27  & 93.1 &  0.68 & 0.02 & 7\\
3457.56 & 0.560  & --0.530  & --0.07  & 33.7 &  0.66 & 0.02 & 7\\
3479.02 & 0.530  & --0.690  & --0.31  & 18.6 &  0.63 & 0.02 & 7\\
3479.39 & 0.710  &   0.170  & --0.47  & 39.3 &  0.63 & 0.01 & 7\\
3499.58 & 0.410  & --0.810  & --0.52  & 13.5 &  0.61 & 0.02 & 7\\
3505.67 & 0.160  & --0.360  & --0.39  & 52.0 &  0.60 & 0.01 & 7\\
3536.94 & 0.360  & --1.310  & --0.22  & 10.5 &  0.57 & 0.03 & 7\\
3551.96 & 0.090  & --0.310  & --0.29  & 64.2 &  0.56 & 0.01 & 8\\
3573.08 & 0.320  & --1.040  & --0.25  & 18.2 &  0.54 & 0.02 & 7\\
3578.23 & 1.220  & --0.610  & --0.20  &  4.9 &  0.54 & 0.05 & 7\\
3630.02 & 0.360  & --1.110  & --0.24  & 14.7 &  0.50 & 0.02 & 7\\
3714.78 & 0.530  & --0.930  & --0.23  & 15.2 &  0.45 & 0.01 & 7\\
3836.77 & 0.560  & --0.060  & --0.39  & 45.3 &  0.39 & 0.01 & 6\\
3998.97 & 0.560  & --0.670  &   0.09  & 39.8 &  0.33 & 0.01 & 7\\
4050.33 & 0.710  & --1.000  & --0.12  & 11.1 &  0.32 & 0.01 & 7\\
4208.98 & 0.710  & --0.460  & --0.13  & 30.7 &  0.29 & 0.01 & 6\\
4317.32 & 0.710  & --1.380  & --0.04  &  6.1 &  0.27 & 0.02 & 6\\\hline
\ion{Nb}{2}, $Z=41$ &  &  &  &  &  \\\hline
3163.40 & 0.376  &   0.260  & --1.63  & 14.9 &  1.47 & 0.06 & 1\\
3215.59 & 0.440  & --0.190  & --1.81  &  3.4 &  1.19 & 0.20 & 6\\\hline
\ion{Mo}{1}, $Z=42$ &  &  &  &  &  \\\hline
3864.10 & 0.000   & --0.010  & --1.00  & 2.7 &  0.38 & 0.07 & 8\\\hline
\ion{Ru}{1}, $Z=44$ &  &  &  &  &  \\\hline
3498.94 & 0.000  & 0.310  & --1.02  & 3.5 &  0.61 & 0.07 & 6\\
3728.03 & 0.000  & 0.270  & --0.81  & 5.3 &  0.44 & 0.03 & 6\\\hline
\ion{Rh}{1}, $Z=45$ &  &  &  &  &  \\\hline
3692.36 & 0.000  & 0.174  & $<$--1.25  & syn &  & & 6\\\hline
\ion{Pd}{1}, $Z=46$ &  &  &  &  &  \\\hline
3404.58 & 0.810  & 0.320  & --1.35  & 6.2* &  0.73 & 0.05 & 6\\\hline
\ion{Ag}{1}, $Z=47$ &  &  &  &  &  \\\hline
3280.68 & 0.000  & --0.050  & --2.03  & 5.1* & 0.97 & 0.10 & 6 \\\hline
\ion{Ba}{2}, $Z=56$ &  &  &  &  &  \\\hline
4554.04 & 0.000  &   0.170  & --1.82  & 96.7 & 0.24 & 0.00 & 6\\
4934.10 & 0.000  & --0.150  & --1.76  & 86.8** & 1.19 & 0.02  & 6\\
5853.70 & 0.604  & --1.010  & --1.58  & 8.4***  & 2.80 & 0.19 & 6\\
6141.70 & 0.704  & --0.070  & --1.67  & 36.1*** & 2.94 & 0.05 & 6\\\hline
\ion{La}{2} $Z=57$ &  &  &  &  &  \\\hline
3794.77 & 0.240  & 0.210  & --2.68  & 3.2* & 0.41 & 0.06 & 7\\
3988.52 & 0.400  & 0.210  & --2.81  & 1.5 & 0.34 & 0.10 & 4\\
3995.75 & 0.170  & --0.060  & --2.81  & 1.6* & 0.34 & 0.09 & 4 \\
4086.71 & 0.000  & --0.070  & --2.54  & 4.7* & 0.31 & 0.03 & 4\\
4123.23 & 0.320  & 0.130  & --2.93  & 1.2* & 0.30 & 0.13 & 4\\\hline
\ion{Ce}{2} $Z=58$ &  &  &  &  &  \\\hline
4222.60 & 0.120  & --0.180  & --2.19  & 1.1 & 0.28 & 0.14 & 6 \\
4523.08 & 0.520  & --0.080  & --1.73  & 1.3* & 0.24 & 0.07 & 10 \\
4539.78 & 0.330  & --0.080  & --2.11  & 1.0 & 0.24 & 0.10 & 10\\
4562.37 & 0.480  & 0.190  & --2.13  & 1.1 & 0.24 & 0.09 &  10\\
4572.28 & 0.680  & 0.290  & --1.95  & 1.1* & 0.24 & 0.09 & 8\\\hline
\ion{Pr}{2} $Z=59$ &  &  &  &  &  \\\hline
4179.40 & 0.200  & 0.480  & --2.22  & 7.3 & 0.29 & 0.02 & 7 \\
4189.48 & 0.370  & 0.380  & --2.18  &  4.0 & 0.29 & 0.03 &8 \\\hline
\ion{Nd}{2} $Z=60$ &  &  &  &  &  \\\hline
3784.25 & 0.380  & 0.150  & --2.28  & 1.6 &  0.41 & 0.13 & 9\\
3826.42 & 0.064  & --0.410  & --2.10  & 1.7 & 0.40 & 0.12 & 9 \\
4061.08 & 0.471  & 0.550  & --2.10  & 4.8 & 0.32 & 0.03 & 9 \\
4232.38 & 0.064  & --0.470  & --1.98  & 2.0* & 0.28 & 0.07 & 9 \\\hline
\ion{Sm}{2} $Z=62$ &  &  &  &  &  \\\hline
4318.94 & 0.280  & --0.270  & --2.52  & 0.8* & 0.27 & 0.21 & 7 \\
4642.23 & 0.380  & --0.520  & --2.31  & 0.6 & 0.23 & 0.18 & 7\\\hline
\ion{Eu}{2} $Z=63$ &  &  &  &  &  \\\hline
3819.67\tablenotemark{a} & 0.000  &  0.510  & --2.96  & syn & &&5\\
4129.70 & 0.000  &  0.220  & --2.89  & 7.7* & 0.30 & 0.02 & 5\\
4205.05\tablenotemark{a} & 0.000  &  0.210  & --2.81  & syn &&& 5\\\hline
\ion{Gd}{2} $Z=64$ &  &  &  &  &  \\\hline
3331.40\tablenotemark{a} & 0.000  & --0.140  & $<$-2.3  & syn & &&7\\
3481.80\tablenotemark{a} & 0.490  & 0.230  & --1.66  & 5.5* & 0.63 & 0.05 & 7\\
3549.37 & 0.240  & 0.260  & --2.92  & 0.7 &  0.56 & 0.85 & 7\\
3768.40 & 0.080  & 0.360  & --2.73  & 2.3 &  0.42 & 0.09 & 7\\\hline
\ion{Dy}{2} $Z=66$ &  &  &  &  &  \\\hline
3460.97 & 0.000  & --0.070  & --2.71  & 1.8  & 0.65 & 0.22 & 7  \\
3531.71 & 0.000  & 0.770  & --2.98  & 6.5 & 0.58 & 0.04 & 8 \\\hline
\ion{Ho}{2} $Z=67$ &  &  &  &  &  \\\hline
3398.94 & 0.000  & 0.410  & $<$--2.20  & syn & &&11 \\\hline
\ion{Er}{2} $Z=68$ &  &  &  &  &  \\\hline
3499.10 & 0.060  & 0.136  & --2.96  & 1.9 &  0.61 & 0.12  & 10\\
3692.65 & 0.050  & 0.138  & --2.63  & 4.3 &  0.46 & 0.06 & 6\\\hline
\ion{Tm}{2} $Z=69$ &  &  &  &  &  \\\hline
3701.36 & 0.000  & --0.540  & $<$--2.8  & syn &&& 8 \\\hline
\ion{Yb}{2} $Z=70$ &  &  &  &  &  \\\hline
3289.37 & 0.000  &  0.020  & --2.99  & 18.2* & 0.95 & 0.03 & 8\\
3694.19 & 0.000  & --0.300  & --2.90  & 13.4* & 0.46 & 0.02 &8\\\hline
\ion{Ir}{1} $Z=77$ &  &  &  &  &  \\\hline
3220.76\tablenotemark{a} & 0.350  & --0.510  & $<$--1.00  & syn &&& 7\\
3800.12 & 0.000  & --1.450  & $<$--1.30  & syn &&& 7 \\\hline
\ion{Th}{2} $Z=90$ &  &  &  &  &  \\\hline
4019.12 & 0.000  & --0.270  & $<$--2.65  & syn & && 7\\\hline
\tableline
\enddata
\tablenotetext{*}{indicate synthesized values calculated for the abundance derived by spectrum synthesis.}
\tablenotetext{**}{indicate the data taken from Honda et al. 2004a}
\tablenotetext{***}{indicate the data taken from Johnson 2002.}
\tablenotetext{a}{$ $Lines that were not used to derive final results.
References.-- 1.Kurucz \& Bell (1995); 2.Westin et al. (2000); 3.Lawler et al. (2001a); 4.Lawler et al. (2001b); 5.Johnson (2002); 6.Hill et al. (2002); 7.Cowan et al. (2002); 8.Sneden et al. (2003); 9.Den Hartog et al. (2003); 10.Johnson et al. (2004); 11.Lawler et al. (2004).}
\end{deluxetable}

\begin{deluxetable}{lcccccccccc}
\tablewidth{0pt}
\tablecaption{ELEMENTAL ABUNDANCES OF HD~88609 and HD~122563 \label{tab:table3}}
\startdata
\tableline
\tableline
& & HD~88609 & & & & HD~122563& & &&\\\hline
species & {\it Z} & log$\epsilon$ & $\sigma$ & [X/Fe] & N & log$\epsilon$ & $\sigma_{\rm random}$ & [X/Fe] & $\sigma_{\rm atm}$\tablenotemark{a} & $\sigma_{\rm n-cap}$\tablenotemark{b} \\\hline
Cu & 29 &  0.66  & 0.20  & --0.48 & 3 & 0.98 & 0.14 & --0.46 & 0.36 & 0.29\\
Zn & 30 &  1.92  & 0.03  & +0.39 & 4 & 2.01 & 0.15 & +0.18 & 0.08 &0.07\\
Sr & 38 & --0.20  & 0.12  & --0.05 & 3 & --0.12 & 0.14 & --0.27 & 0.14, 0.34 & 0.10, 0.27\\
Y  & 39 & --0.98  & 0.10  & --0.12 & 15 & --0.93 & 0.09 & --0.37 &0.20 & 0.12\\
Zr & 40 & --0.24  & 0.16  & +0.24 & 17 & --0.28 & 0.16 & --0.10 & 0.15 & 0.06\\
Nb & 41 & --1.72  & 0.12  & --0.07 & 2 & --1.48 & 0.14 & --0.13 & 0.14 & 0.06\\
Mo & 42 & --1.00  & 0.12  & +0.15 & 1 & --0.87 & 0.14 & --0.02 & 0.19 & 0.12\\
Ru & 44 & --0.91  & 0.12  & +0.32 & 2 & --0.86 & 0.14 & 0.07 & 0.20 & 0.12\\
Rh & 45 & $<$--1.25  &       & $<$+0.70 & 1 & $<$--1.20 && $<$+0.45\\
Pd & 46 & --1.35  & 0.12  & +0.03 & 1 & --1.31 & 0.14 & --0.23 & 0.20 & 0.12\\
Ag & 47 & --2.03  & 0.12  & +0.10 & 1 & --1.88 & 0.14 & --0.05 & 0.21 & 0.13\\
Ba & 56 & --1.71  & 0.10  & --0.81 & 4 & --1.65 & 0.12 & --1.05 & 0.15 & 0.05\\
La & 57 & --2.75  & 0.15  & --0.81 & 5 & --2.60 & 0.16 & --0.96 & 0.13 & 0.08\\
Ce & 58 & --2.02  & 0.18  & --0.53 & 5 & --1.91 & 0.17 & --0.72 & 0.12 & 0.07\\
Pr & 59 & --2.20  & 0.12  & +0.14 & 2 & --2.15 & 0.14 & --0.09 & 0.14 & 0.06\\
Nd & 60 & --2.11  & 0.12  & --0.49 & 4 & --2.01 & 0.16 & --0.69 & 0.13 & 0.07\\
Sm & 62 & --2.41  & 0.12  & --0.35 & 2 & --2.16 & 0.14 & --0.40 & 0.13 & 0.07\\
Eu & 63 & --2.89  & 0.12  & --0.33 & 1 & --2.77 & 0.14 & --0.52 & 0.14 & 0.08\\
Gd & 64 & --2.83  & 0.12  & --0.88 & 2 & --2.44 & 0.14 & --0.76 & 0.15 & 0.07\\
Dy & 66 & --2.85  & 0.20  & --0.92 & 2 & --2.62 & 0.14 & --0.99 & 0.14 & 0.08\\
Ho & 67 & $<$--2.20  &       & $<$+0.36 & 1 & $<$--2.00 && $<$+0.26\\
Er & 68 & --2.79  & 0.20  & --0.65 & 2 & --2.66 & 0.14 & --0.82 & 0.14 & 0.08\\
Tm & 69 & $<$--2.80  &       & $<$+0.27 & 1 & $<$--3.00 && $<$--0.23\\
Yb & 70 & --2.94  & 0.12  & --0.95 & 2 &  --2.78 & 0.14 &  --1.09 & 0.14 & 0.06\\
Ir & 77 & $<$--1.30  &       & $<$+0.39 & 2 & $<$--1.60 && $<$--0.21\\
Th & 90 & $<$--2.65  &       & $<$+0.36 & 1 & $<$--3.05 && $<$--0.34\\
\tableline
\enddata
\tablenotetext{a}{$\sigma_{\rm atm}$ indicates the root sum square of the uncertainties contributed by the four atmospheric parameters.}
\tablenotetext{b}{$\sigma_{\rm_n-cap}$ indicates the fluctuation of the quantity of the change for each elements which are the differences from the average of quantity with atmospheric parameter changed in HD~122563.}
\end{deluxetable}

\clearpage

\clearpage

\clearpage

\clearpage

\clearpage

\clearpage

%

\end{document}